\documentclass[preprint,numbers,10pt]{sigplanconf}

\usepackage{amsmath}
\usepackage{amssymb}
\usepackage{url}		
\usepackage{listings}		
\usepackage{enumitem}		
\usepackage{color}
\usepackage[pdftex,dvipsnames]{xcolor}
\usepackage{tikz}
\usepackage{multicol}
\usepackage{multirow}
\usepackage{hhline}
\usepackage[linesnumbered, lined, boxed]{algorithm2e}
\usepackage{lipsum}		
\usepackage{xargs}		
\usepackage[colorinlistoftodos,prependcaption,textsize=tiny]{todonotes}

\usepackage{microtype}
\usepackage{vwcol}

\definecolor{colKeys}{rgb}{0,0,1}
\definecolor{colIdentifier}{rgb}{0,0,0}
\definecolor{colComments}{rgb}{0,0.5,1}
\definecolor{colString}{rgb}{0.6,0.1,0.1}
\newcommand{\T}{{\rule{0pt}{2.2ex}}}       
\newcommand{\B}{{\rule[-0.8ex]{0pt}{0pt}}} 

\newcommandx{\unsure}[2][1=]{\todo[linecolor=red,backgroundcolor=red!25,bordercolor=red,#1]{#2}}
\newcommandx{\change}[2][1=]{\todo[linecolor=blue,backgroundcolor=blue!25,bordercolor=blue,#1]{#2}}
\newcommandx{\info}[2][1=]{\todo[linecolor=OliveGreen,backgroundcolor=OliveGreen!25,bordercolor=OliveGreen,#1]{#2}}
\newcommandx{\improvement}[2][1=]{\todo[linecolor=Plum,backgroundcolor=Plum!25,bordercolor=Plum,#1]{#2}}
\newcommandx{\thiswillnotshow}[2][1=]{\todo[disable,#1]{#2}}

\usetikzlibrary{snakes,arrows,shapes,positioning,fit}
\usetikzlibrary{decorations.pathmorphing}
\pgfdeclarelayer{background}
\pgfdeclarelayer{foreground}
\pgfsetlayers{background,main,foreground}

\newcommand{\doi}[1]{doi:~\href{http://dx.doi.org/#1}{\Hurl{#1}}}   

\newtheorem{theorem}{Theorem}
\newtheorem{lemma}[theorem]{Lemma}
\newtheorem{definition}[theorem]{Definition}

\newtheorem{propal}[theorem]{Proposition}

\begin{document}

\setlength{\pdfpageheight}{\paperheight}
\setlength{\pdfpagewidth}{\paperwidth}

\conferenceinfo{CONF 'yy}{Month d--d, 20yy, City, ST, Country}
\copyrightyear{20yy}
\copyrightdata{978-1-nnnn-nnnn-n/yy/mm}
\copyrightdoi{nnnnnnn.nnnnnnn}



\title{Hybrid Static/Dynamic Schedules for Tiled Polyhedral Programs}

\authorinfo{Tian Jin}
{\makebox{Computer Science Department} \\
  \makebox{Haverfort College} \\
  \makebox{\texttt{tjin@haverford.edu}}}

\authorinfo{Nirmal Prajapati, Waruna Ranasinghe, Guillaume Iooss, Yun Zou,
  Sanjay Rajopadhye}
{\makebox{Computer Science Department} \\
  \makebox{Colorado State University} \\
  \makebox{\texttt{[First.Last]@colostate.edu}}}

\authorinfo{David Wonnacott}
{\makebox{Computer Science Department} \\
 \makebox{Haverfort College} \\
   \makebox{\texttt{davew@cs.haverford.edu}}}

\maketitle

\begin{abstract}

  Polyhedral compilers perform optimizations such as tiling and
  parallelization; when doing both, they usually generate code that executes
  ``barrier-synchronized wavefronts'' of tiles.  We present a system to
  express and generate code for hybrid schedules, where some constraints are
  automatically satisfied through the structure of the code, and the remainder
  are dynamically enforced at run-time with data flow mechanisms.  We prove
  bounds on the added overheads that are better, by at least one polynomial
  degree, than those of previous techniques.

  We propose a generic mechanism to implement the needed synchronization, and
  show it can be easily realized for a variety of targets: OpenMP, Pthreads,
  GPU (CUDA or OpenCL) code, languages like X10, Habanero, Cilk, as well as
  data flow platforms like DAGuE, and OpenStream and MPI.  We also provide a
  simple concrete implementation that works without the need of any
  sophisticated run-time mechanism.

  Our experiments show our simple implementation to be competitive or better
  than the \textit{wavefront-synchronized code} generated by other systems.
  We also show how the proposed mechanism can achieve 24\% to 70\% reduction
  in energy.
\end{abstract}

%
%

\section{Introduction}
\label{intro}

The ongoing evolution and increasing complexity of modern computer architecture
creates new challenges to the goal of tuning software for optimal performance.
For example, the exact time for any event, across multiple levels of the
compute/memory hierarchy, is increasingly unpredictable because of latency and
bandwith differences in accesses to various levels of caches, and instruction
pipelining and reordering, performed in the microarchitecture.  Indeed, most
machines can be abstracted as coupled microarchitectural modules interacting
more like ``data-flow engines,'' rather than simple ``data-paths'' controlled
by finite state machines.  


To meet these challenges 
we must create tools applicable to 
varied target architectures.  The development of the \emph{polyhedral
  model}~\cite{sanjay-fst-tcs, quinton-jvsp89, feautrier91, feautrier92a,
  Pugh:92} enables analyses like automatic parallelization on a significant
class of programs, and transformations like loop tiling~\cite{IT:88, WL:91} to
be applied at a scope where it is most valuable~\cite{Wonnacott:02}.  While
challenges remain in expanding the domain of the polyhedral model, it is
already widely used in research~\cite{uday-pldi08, pouchet-etal-sc10,
  YBGIKPSZR:12} and commercial~\cite{MLVWB:09} compilers.

State of the art polyhedral compilers first choose a \emph{schedule} that
identifies the \emph{potentially tilable} dimensions, and also the
\emph{inherently sequential} dimensions within the iteration spaces of the
program.  Subsequently they generate code that visits the inherently
sequential dimensions in serial order, tiles the tilable dimensions, and
executes the tiles in parallel, typically via an outer sequential loop that
enumerates ``wavefronts'' of tiles, such that every tile depends only on tiles
from ``strictly previous'' wavefronts.  A sequential \texttt{for} loop around
an OpenMP \texttt{parallel for} loop, together with a barrier between
consecutive wavefronts, provides a convenient and portable mechanism to
express this execution.

As many authors note, such static control structures have a number of
drawbacks~\cite{KPPGCS2015DataflowTiled, DAGuE,
  Baskaran:2009:CDS:1504176.1504209, BBDHD2012EuroPar,
  dathathri-etal-topc2016, ZR:15}.  First, they induce unnecessary
synchronization---any tile of wavefront~$w$ \emph{must wait} for \emph{all}
tiles of wavefront~$w-1$.
Second, such over-constrained schedules may prevent other optimizations like
loop fusion~\cite{KPPGCS2015DataflowTiled}.  Third, it suffers what we call
the \emph{affinity problem}:
data needed for a tile may not reside in that (or any) core's cache at the
start of tile execution---other tiles in the previous wavefront may have
evicted it.  This has a cost on both time and energy, although ``multipass
wavefront schedules'' can reduce it~\cite{ZR:15}.

To address these drawbacks, many authors suggest compiling even ``regular''
programs such as dense linear algebra or polyhedral loops to a data-flow
runtime/library on shared memory
platforms~\cite{Baskaran:2009:CDS:1504176.1504209, KPPGCS2015DataflowTiled,
  BBDHD2012EuroPar} or distributed memory machines~\cite{DAGuE, Codelets,
  dathathri-etal-topc2016} or hybrid systems~\cite{dathathri-etal-topc2016}.
Similar solutions are also available on accelerators~\cite{waruna-ms-thesis,
  DAGuE, dathathri-etal-topc2016}.
Regardless of the specific target language/platform, we call this
\emph{data-flow runtime synchronization}, since many are inspired by earlier
work on general purpose data-flow computing systems~\cite{sato-etal-isca92,
  theobald99-earth}.


However, data-flow runtime systems incur overheads in: task synchronization,
memory for managing task state, and most importantly, task creation/spawning.
All systems synchronize the dependences of each tile, and some have a memory
overhead that is directly proportional to the total number of tiles.  Some
even construct the entire task graph at compile time.  Significantly,
\emph{all existing data-flow systems for polyhedral programs create and launch
  one task per tile}.  This not only contributes to inefficiencies in context
switching and scheduling overhead, but also suffers from the affinity problem:
the time/processor when/where a tile is eventually scheduled is decided by the
data-flow run-time.

In this paper, we show that for \emph{polyhedral programs}, such approaches
are overkill, and develop a very simple, domain-specific solution.  Our
granularity of task creation/spawning is not a tile, but what we call a
\emph{virtual processor}: a sequence or ``slice'' of tiles.  This reduces the
number of tasks created by a polynomial degree and the synchronization tests
by a constant factor (see the asymptotic complexity analyses in
Sec.~\ref{sec:complexity}).  We use a partial order view of polyhedral
time~\cite{sanjay-ppopp2013} to define \emph{hybrid static-dynamic} schedules,
provide their legality conditions, and show how this leads to an algorithm to
separate dependences guaranteed to be \emph{statically satisfied} from those
which must be satisfied via a \emph{run-time check} (see Sec.~\ref{sec:hsd}).
We then formalize the precedence obligations of these checks, and develop a
\emph{self-scheudling run-time} mechanism (in Sec.~\ref{sec:runtime}) that
guarantees these dependences without any of the overheads of a full fledged
data-flow run-time system.

Our solution is also generic, and leads to a single code generator framework
to produce code with target-agnostic synchronization ``stubs,'' that can be
filled in with primitives for a variety of targets, such as OpenMP,
C+Pthreads~\cite{Pthreads}, CUDA\cite{CUDA}, X10~\cite{X10},
Habanero~\cite{Habanero}, Cilk++~\cite{Leiserson2009},
OpenStream~\cite{OpenStream}, and under simple instances, also MPI.
Furthermore, it also allows us to cleanly isolate issues like deadlock that
often depend on some specific properties of the target.

We provide an implementation of the synchronization stubs for two platforms:
OpenMP and C+Pthreads.  The lightweight runtime leads to significant
performance gains in both time and energy (see Sec.~\ref{sec:experiments}).

\section{Background}	
\label{sec:background}

We now describe the background on polyhedral compilation needed to make the
paper self contained.  Some of our explanations are (deliberately) simplistic
in order to convey the main intuitions concisely.

\begin{figure}[tb]
  {\small
\begin{lstlisting}
// H initialized to 0
  for (i = 0; i < M; i++)
    for (j = 0; j < N; j++)
S:    H[j] = foo(H[j], H[j-1]);
  return H[N-1]
\end{lstlisting}
}
\caption{\small{REX (Running EXample)}}
\label{fig:motiv}
\end{figure}

\begin{figure}[t]
  \centering 
  \includegraphics[scale=0.4]{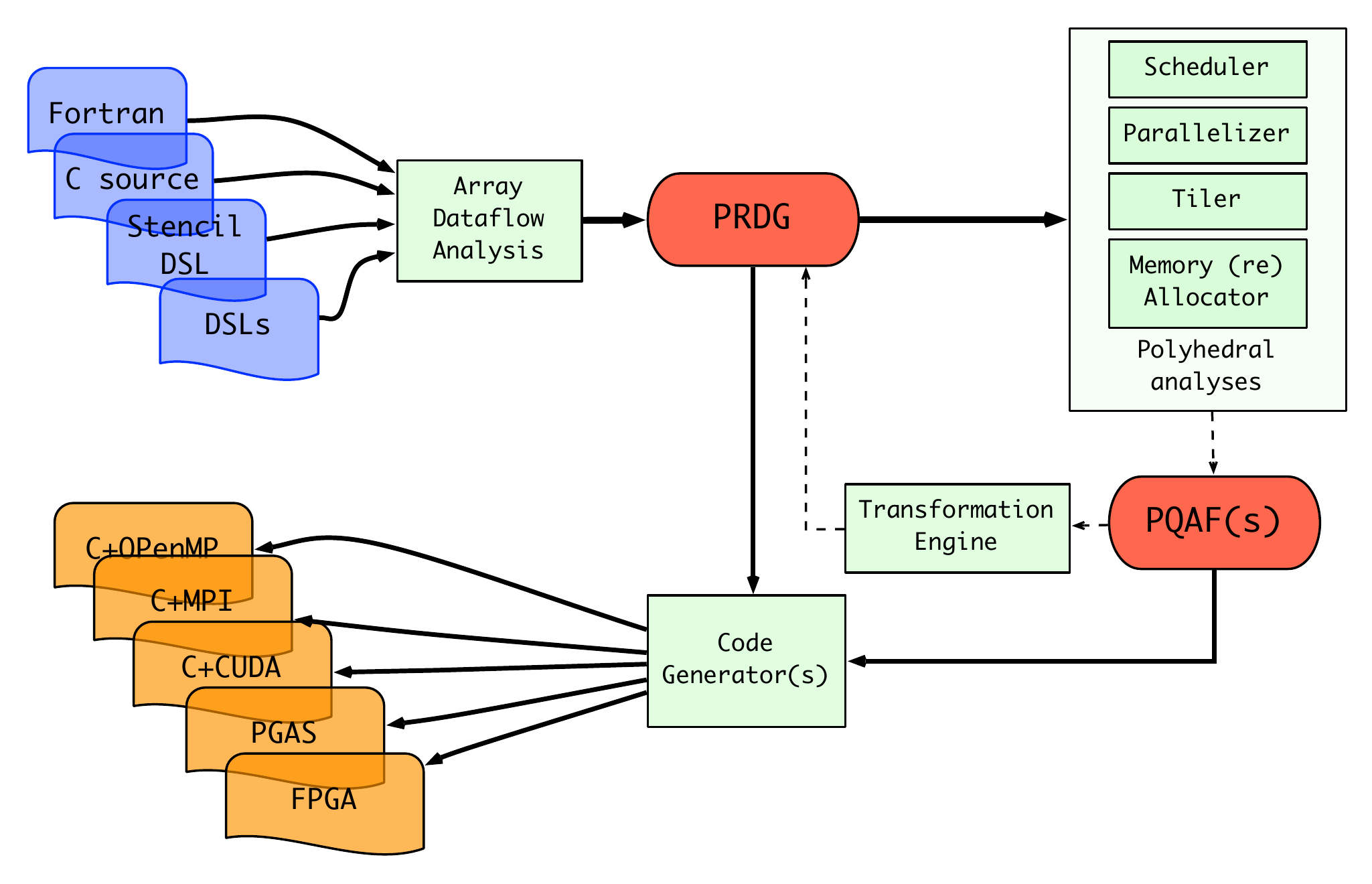}
  \caption{\small{Polyhedral Compilation: the Polyhedral Reduced Dependence
      (hyper) Graph (PRDG) serves as the intermediate representation, with
      Piecewise Quasi-Affine Functions (PQAFs) describing transformations of
      this representation.
	}}
  \label{fig:compiler}
\end{figure}

The polyhedral framework lets compilers reason about loop nests in terms of
{\em executions} or \emph{instances} of statements rather than statements
themselves.  Consider the code of Figure~\ref{fig:motiv}, which we will use as
a simple running example.  It implements the following equation ($H[M-1,N-1]$
is the desired output):
\begin{equation}
  H[x,y] = \left\{\begin{array}{ll}
      0 & \mathit{if}\ \  x = 0 \lor y = 0 \\
      \mathit{foo}(H[x-1,y], H[x,y-1]) & \mathit{otherwise}
    \end{array} \right.
\end{equation}


The polyhedral model lets a compiler reason about
an $M\times N$ set of executions of that statement (labelled S),
which we could express in
the notation of \emph{isl}~\cite{Verdoolaege:10} as
\begin{equation*}
[N, M] \rightarrow \{ (i, j) \mid 1 \leq i < M \land 1 \leq j < N \}
\end{equation*}
This representation lets the compiler reason about iteration spaces of unknown
size by describing potentially infinite sets of integer points with a finite
set of affine equality and inequality constraints on integer variables
(i.e., they reason about sets of integer points inside a potentially unbounded polyhedron).
While decision algorithms for this domain have high complexity,
they typically perform well in this context due to the simplicity
of the constraints involved~\cite{pugh-toplas98}.

Array dataflow analysis~\cite{feautrier91,PW:92} (ADA) lets the compiler extract the flow
of information in the code,
for example recreating the $otherwise$ clause of
the original equations at the top of Figure~\ref{fig:motiv}
by identifying
the sources of the two values arriving at iteration $(i, j)$ of Statement $S$
from earlier executions of $S$:
\begin{equation*}
  \begin{split}
    & \{ (i, j) \rightarrow (i-1, j) \mid 2 \leq i \leq M \land 1 \leq j \leq
    N \} \\
    \cup \ & \{ (i, j) \rightarrow (i, j-1) \mid 1 \leq i \leq M \land 2 \leq
    j \leq N \}
  \end{split}
\end{equation*}

This representation provides the information needed to test the legality of
any reordering of iterations.
We can furthermore use the polyhedral framework to find the flow of
information among tiles,
and use that to determine the possible legal execution orderings of tiles.

\begin{figure*}[t]
\begin{vwcol}[widths={0.35,0.65}, rule=0pt]
	\includegraphics[scale=0.55]{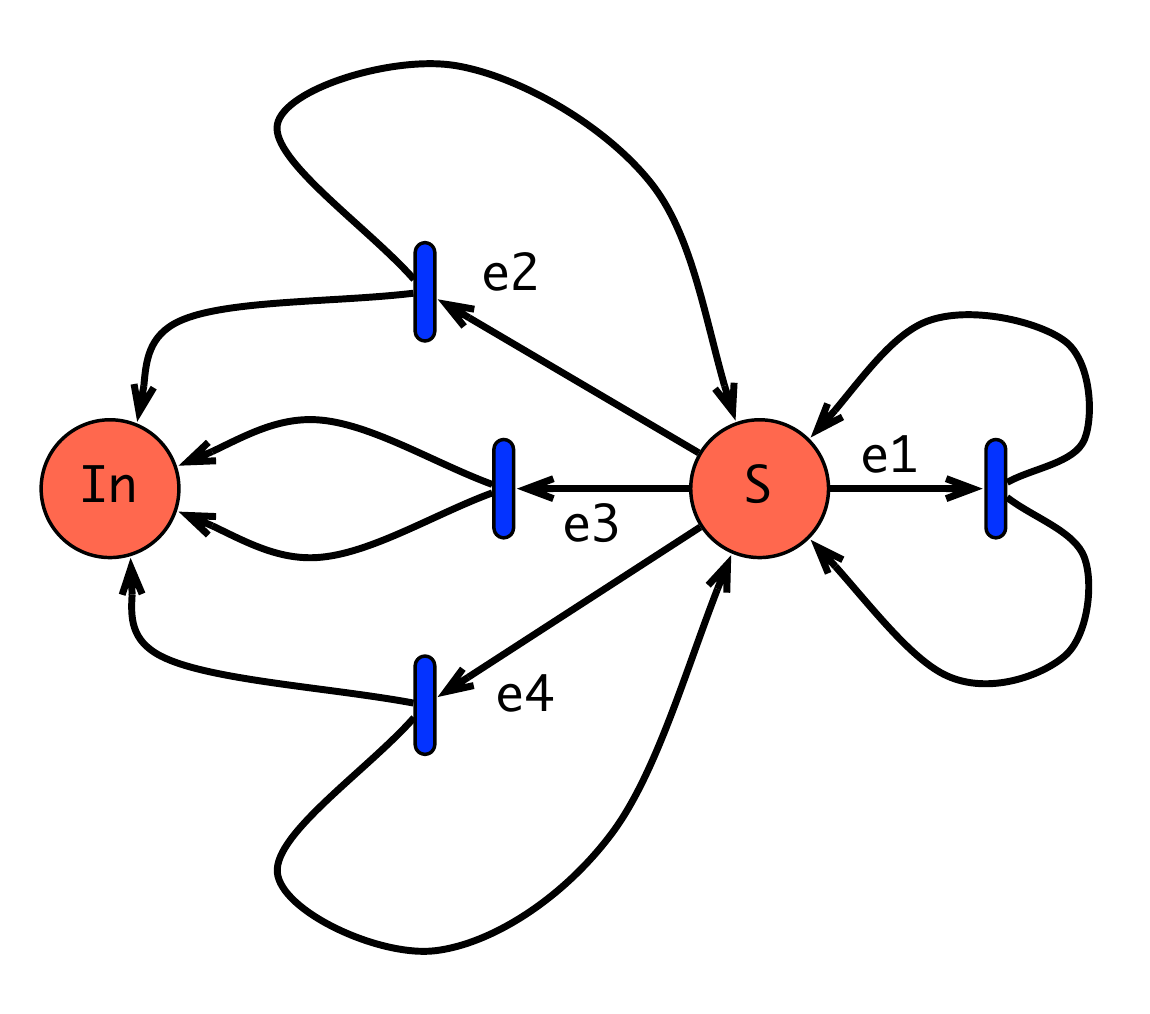}
	
	\columnbreak
	\small
	\noindent \textbf{Original PRDG:}
	\begin{itemize}
		\item S: $\{i,j ~|~ 1\leq i \leq M \land 1\leq j\leq N \}$
		\item In: $\{i,j ~|~ i=0 \land 1\leq j\leq N\} \cup \{i,j ~|~ j=0 \land 1\leq i \leq M \}$
		\item e1 ($i > 1 \land j > 1$):
			 $(i,j \mapsto S(i-1,j))$, $(i,j \mapsto S(i,j-1))$
		\item e2 ($i > 1 \land j=1$):
			 $(i,j \mapsto S(i-1,j))$, $(i,j \mapsto In(i,0))$
		\item e3 ($i = 1 \land j = 1$):
			 $(i,j \mapsto In(0,j))$, $(i,j \mapsto In(i,0))$
		\item e4 ($i = 1 \land j > 1$):
			 $(i,j \mapsto In(0,j))$, $(i,j \mapsto S(i,j-1))$
	\end{itemize}
	\textbf{Tiled PRDG:} the nodes and edges are the same
	\begin{itemize}
		\item S: $\{i_b,j_b ~|~ 0\leq i_b \leq M_b \land 0\leq j_b\leq N_b \}$
		\item In: $\{i_b,j_b ~|~ i_b=0 \land 1\leq j_b\leq N_b\} \cup \{i_b,j_b ~|~ j_b=0 \land 1\leq i_b \leq M_b \}$
		\item e1 ($i_b > 0 \land j_b > 0$):
			$(i_b,j_b \mapsto S(i_b-1,j_b))$, $(i_b,j_b \mapsto S(i_b,j_b-1))$
		\item e2 ($i_b > 0 \land j_b=0$):
			$(i_b,j_b \mapsto S(i_b-1,j_b))$, $(i_b,j_b \mapsto In(i_b,0))$
		\item e3 ($i_b = 0 \land j_b = 0$):
			 $(i_b,j_b \mapsto In(0,j_b))$, $(i_b,j_b \mapsto In(i_b,0))$
		\item e4 ($i_b = 0 \land j_b > 0$):
			 $(i_b,j_b \mapsto In(0,j_b))$, $(i_b,j_b \mapsto S(i_b,j_b-1))$
	\end{itemize}
\end{vwcol} 
  \centering
  \caption{\small{The PRDG for REX,
      with original (inter-statement) and some inter-tile dataflow edges.
	}}
  \label{fig:prdg}
\end{figure*}

Figure~\ref{fig:compiler} illustrates the flow of a typical polyhedral
compiler.  The input program may be in any standard imperative language,
possibly with annotations as to which part of the code is to be analyzed using
polyhedral techniques.
The front end produces descriptions of the iteration spaces and the
inter-iteration dataflow (as discussed above) among all statements; depending
on the set of transformations to be applied, it may also record the memory
address written by each iteration.  These data structures are collectively
known as the \emph{Polyhedral Reduced Dependence (hyper) Graph} (PRDG), and
serve as the intermediate representation for a polyhedral compiler.
Additional analysis steps (or programmer direction) are used to define program
transformations as a set of functions on the iteration space.  Like the
descriptions of iteration spaces and dataflow, these are represented via
affine constraints on integer variables, or the union of several such
representations, known as \emph{(Piece-wise) Quasi-Affine Functions} (PQAF's).
The PQAF's for the program transformations can be fed (along with the PRDG)
into a code generator~\cite{Verdoolaege:10} or used iteratively to transform
the PRDG itself.

\subsection{Polyhedral Reduced Dependence (hyper) Graph}
\label{sec:prdg}

Figure~\ref{fig:prdg} gives the PRDG for REX.  Nodes correspond to statements
(with corresponding iteration spaces shown to the right).  Although the loop
contains only one statement ($S$), a pseudo-statement $\mathit{In}$ is
automatically generated to define the source of live-in array elements.  PRDG
hyper-edges are used to describe dependences; there are several variants on
this technique.  Here, we use one hyper-edge for each region within the
iteration space, and our hyper-edges have one source node and possibly
multiple destination nodes.  For each destination node $Y$ of a hyper-edge we
identify an ``input-output pair'' with the 4-tuple $\langle X,Y,\mathcal{D},f
\rangle$, where $X$ is the source of the hyper-edge, $f$ and $\mathcal{D}$ are
respectively the value and the constraints of one of the case derived through
the ADA.

For example, for REX, the $S$ node depends on the $\mathit{In}$ node for both
its inputs in the first iterations of the {\tt i} and {\tt j} loops.
Similarly there are three other cases:
\begin{itemize}\itemsep 0mm
\item for points in the interior, i.e., $(i,j)>1$, the two dependences are
  from earlier iterations of statement $S$,
  specifically $\langle i, j-1\rangle$, and $\langle i-1, j\rangle$
  (this is Edge e1 in Figure~\ref{fig:prdg}),
\item for $i>1, j=1$ (Edge e2), the first reference is live-in to the loop
  (from $\bot$, shown as $\mathit{In}$ in our PRDG), and the second from
  $\langle i-1, j\rangle$,
\item for $i=j=1$ (Edge e3), both references are $\bot$, and
\item for $i=1, j>1$ (Edge e4), the first reference is $\langle i, j-1\rangle$ and the
  second one is to $\bot$.
\end{itemize}

\subsection{Tiling and the Tiled PRDG}
\label{sec:hprdg}

In polyhedral compilation, tiling is possibly \emph{the most} critical
transformation.  It groups computations/iterations into \emph{tiles}, which
are executed \emph{atomically}.  There is a huge amount of literature on the
subject of tiling, including the feasible space of ``legal'' tiling
hyperplanes, the choice of ``good'' tiling hyperplanes and tile sizes that
seek to optimize a wide range of cost functions, addressing locality,
coarse-grain and/or fine-grain parallelism, etc.  This is completely
orthogonal to our work.  We assume that a tiling has been chosen. We present a
simplified description of how tiling affects the program representation.

Tiles can be of various shapes (e.g., parallelogram, hexagon) and sizes
important attribute of tiling is the tile size, for which there are two main
choices: fixed, or parametric.  With fixed-size tiling, the resulting program
remains polyhedral and can be further analyzed and transformed using
polyhedral techniques.  However, it means that when the space of tile sizes is
to be explored, say by auto-tuners, the code generation and compilation must
be done repeatedly.  Moreover, it precludes delayed binding of tile sizes.
Parametric tiling, on the other hand allows such delayed binding of tile
sizes, but since it is inherently a nonlinear transformation, the program is
no longer polyhedral, and cannot be further analyzed, transformed or
optimized.  Recently, Iooss et al.~\cite{iooss:15} introduced
\emph{monoparametric tiling} that provides a compromise: it remains a
polyhedral transformation, but allows only a single tile size parameter.  In
this paper, our goal is to further analyze and/or transform the program after
tiling, so we use fixed-size or monoparametric tiling.

Tiling introduces a new set of indices called \emph{tile indices} which
identifies a tile in the computation.  Depending on the number of original
dimensions that are tiled, the total number of indices could increase by up to
a factor of~2.

In this paper, we are only interested in the inter-tile dependences.  By
analyzing these dependences (provided the program remains polyhedral, i.e., if
we use fixed-size or monoparametric tiling), we can build the \emph{tiled
  PRDG} which considers the tiles of a program at points in an iteration
space, and the dependences across them.  Here, all the domains associated with
the nodes and affine functions associated with the hyper-edges are
\emph{projections on the tile indices} of the original domains and
dependences.  Thus, the tiled PRDG will be very similar to the original PRDG,
except that the labels on the hyper-edges now represent inter-tile
dependences, and the domains and the functions of this representation.
Indeed, for REX, the PRDG is (coincidentally) isomorphic to the original PRDG
(see Figure~\ref{fig:prdg}).

\section{Hybrid Schedules}
\label{sec:hsd}

For the purpose of this paper, we assume that the program has been
tiled,
  and the specific problem we
address is its subsequent ``coarse-grain'' parallelization.  We start with the
program representation in the form of a Tiled PRDG, where each node represents
a ``sort'' or ``signature'' of tiles, the domain of the node defines the set
of instances of tiles of that particular sort, and hyper-edges define
\emph{inter-tile} dependences.  We achieve ``coarse-grain'' parallelization
using a hybrid schedule that is partially static and partially dynamic.

\begin{definition}
  For any node $X$ in the PRDG, an $m$-dimensional \emph{schedule} $\theta_X$
  is an affine function that maps every point to an integer vector $\vec{t}
  \in {\cal Z}^m$.
\end{definition}

There is a large body of work on how such schedules can be determined,
including optimality conditions under a range of cost
models~\cite{sanjay-fst-tcs, quinton-sanjay-tf, quinton-jvsp89, feautrier92a,
  DKR-ppl91, DR-tpds94, DR-jpdc95}.  When extended to multi-dimensional
time~\cite{feautrier92b}, precedence is the lexicographic order: an
$m$-dimensional time stamp, $t_1$ happens before another one, $t_2$ iff $t_1$
precedes $t_2$ in the lexicographic order.  It is important to note that in
this early work time is viewed as a \emph{total order}: given two distinct
time vectors, it is always possible to determine which one occurs before the
other.  Such schedules are routinely used, albeit in an ad-hoc manner, by
polyhedral compilation tools to generate code where the execution order is
actually a \emph{partial order} (e.g., ``outer-parallel''
\textbf{\texttt{for}} loops in OpenMP).  Recent work by Verdoolaege et.\
al~\cite{sven-impact14} and Yuki et.\ al~\cite{sanjay-ppopp2013} formalizes
partial order schedules, and we use a similar notion here.

\begin{definition}
  For any node $X$ in the PRDG, a \textbf{hybrid static-dynamic} (HSD)
  \textbf{schedule}, $\theta_X$ is an affine function that maps every point in
  the domain associated with the node to an integer $n$-vector $\vec{t}$,
  called the \textbf{space-time}, or the \textbf{processor-time} vector.  An
  HSD schedule has the following properties.
  \begin{itemize}
  \item For some integer $k, ~ 0\leq k \leq n$ the first $k$ dimensions of
    $\vec{t}$ are called the (virtual) processor/space dimensions, and the
    remaining $n-k$ dimensions, are called the (local) time dimensions.  The
    corresponding subvectors of $\vec{t}$ are $\vec{t}_p$, $\vec{t}_t$ and the
    function $\theta_X$ has two components: $\pi_X$ and $\tau_X$: $\vec{t}_p =
    \pi_X(z)$, and $\vec{t}_t = \tau_X(z)$.
  \item Space-time vectors form a partial order: $\vec{t}$ happens before
    $\vec{t'}$ iff $\vec{t}_p = \vec{t}_p'$, and $\vec{t}_t$ precedes
    $\vec{t}'_t$ in the lexicographic order; if $\vec{t}_p \neq \vec{t}_p'$,
    the two are incomparable.
  \item $\theta_X$ is a bijection.
  \end{itemize}
\end{definition}

\begin{definition}
  An HSD schedule for a program is a set of HSD schedules, one for each node
  in its PRDG, that maps iteration points to a \textbf{common} space-time,
  i.e., with the property that $n$ and $k$ are identical for all variables.
\end{definition}
\subsection{Partial Legality}

We now describe (some of) the legality conditions that HSD schedules must
satisfy.  Informally, whenever $X[z]$ depends on $Y[f(z)]$, and the two
instances are both assigned the same processor, the respective time steps of
these two points must satisfy (strict) lexicographic precedence, so as to
ensure that the ``producer happens before the consumer.''

\begin{definition}
  An HSD schedule for a PRDG, is said to \textbf{partially respect} an
  input-output pair $\langle X, Y, D, f\rangle$, iff
  \[ \forall z \in D, \pi_X(z) = \pi_Y(f(z)) \Rightarrow \tau_X(z) \succ
  \tau_Y(f(z))
  \]

  An HSD schedule for a PRDG partially respects a hyper-edge of the PRDG iff it
  partially respects all the input-output pairs of the hyper-edge.

  An HSD schedule for a PRDG is partially legal iff it partially respects all
  the hyper edges of the PRDG.
\end{definition}

Note that our definition of legality, and respecting dependences/hyper edges is
silent about what happens when the producers and consumers are mapped to
distinct (virtual) processors.  This is because, time stamps where this does
not hold, are not comparable, and require a separate mechanism to enforce the
legality of the final program.  Regardless of how this mechanism is
implemented, it does not need to satisfy all instances of all program all
dependences, but rather a subset, that we call the \emph{residual
  dependences}.


\subsection{Residual PRDG and full Legality}

Algorithm~\ref{alg:residual_prdg}, describes how to isolate the residual
dependences, in the form of a transformed PRDG, called the Residual PRDG
(RPRDG).

\begin{algorithm}[h] 
  \label{alg:residual_prdg}
  \KwIn{PRDG $\cal H$, HSD Schedule (aka Target Mapping) for each node in
    $\cal H$}
  \KwOut{Residual PRDG $\cal H'$}
  \ForEach {hyper-edge $h$ of $\cal H$ with source $X$ and domain $\cal D$}{
    \ForEach {destination, $Y$ of $h$ with function $f$}{
      ${\cal C} = \{z~|~\leftarrow$ $\pi_X(z)\neq \pi_Y(f(z)) \}$ \;
      ${\cal D} \leftarrow {\cal D} \cap {\cal C}$ \;
      \If{${\cal D}$ is empty}{
        delete hyper-edge $h$\;
        break out to next $h$\;}
    }
  }
  \Return {\cal H}
\caption{Constructing the Residual PRDG}
\end{algorithm}

For the REX in Fig~\ref{fig:prdg}, the RPRDG will have same two nodes with two
edges from node $S$ to itself. e1 ($i_b > 0 \land j_b > 0$): $(i_b,j_b \mapsto
S(i_b-1,j_b))$ and e2 ($i_b > 0 \land j_b=0$): $(i_b,j_b \mapsto
S(i_b-1,j_b))$.

There is a plethora of languages and associated run-time
systems~\cite{Leiserson2009, CnC, TBB, Pthreads, UPC, Titanium, X10, Chapel,
  Habanero, Codelets, DAGuE} to ensure dependences are dynamically satisfied.
However, because we are dealing with polyhedral programs, a very small number
of simple abstractions are sufficient.


Before proceeding further, we perform a simple transformation on the RPRDG
(following the dotted lines in Fig.~\ref{fig:compiler}).  We use our HSD
schedule functions $\theta_X, \theta_Y, \ldots$ (recall that they are all
bijections) to ``rename/reindex'' the program so that all nodes, domains, and
dependences are brought into a common set of space-time coordinates, $\vec {z}
= \langle\vec{p}, \vec {t} \rangle$.  the for any node, $X$, ${\cal D}_X$
denotes the set of space-time coordinates where tiles with signature $X$ are
to be executed.  Consider an input-output pair $\langle {\cal D}, X, Y, f
\rangle$, and let us separate out the space and time components.  Let or
$\vec{z} = \left(\begin{array}{c} \vec{p} \\ \vec{t} \end{array}\right) \in
{\cal D}$, and let $f(\vec{z}) = \left(\begin{array}{c} f_p(\vec{z}) \\
    f_t(\vec{z}) \end{array}
\right) = \left(\begin{array}{c} \pi(\vec{p}, \vec{t}) \\
    \tau(\vec{p}, \vec{t}) \end{array} \right)$.

The dynamic dependences obligations of each input-output pair in the RPRDG,
are as follows.

\begin{propal} \label{prop:dep} The tile with signature $X$ at $\vec {z} =
  \langle\vec{p}, \vec {t} \rangle$ cannot be executed until the following
  constraint holds: \\
  If $\vec{z}\in {\cal D}$ then the tile with signature $Y$ at space-time
  coordinates $\langle \pi(\vec{p}, \vec{t}), \tau(\vec{p}, \vec{t}) \rangle$
  must have completed execution.
\end{propal}

\section{A Self-Scheduling Runtime}
\label{sec:runtime}

Our strategy for dynamically satisfying residual dependences is simple and
consists of the following elements.  We maintain a global, shared data
structure (an array) called $\mathtt{\mathbf{State}}_X$ that stores the status
of tiles with signature $X$ on each (virtual) processor.  The $\vec{t}_p$-th
entry in this array is an $(n-k)$-tuple of integers, representing the most
recent time-step that processor $\vec{t}_p$ has completed.

Each processor executes a \emph{sequential} program that visits tiles
allocated to it in the lexicographic order of its \emph{local time stamp}.  At
each iteration, it executes the following steps:
\begin{itemize}\itemsep 0mm
\item Call a meta-function $\mathtt{\mathbf{acquire(X, \vec{p},\vec{t})}}$, to
  ensure that all its residual dependences have been satisfied (as per
  Prop.~\ref{prop:dep}).
\item Execute $\mathtt{\mathbf{Tile_X (\vec{p},\vec{t})}}$
\item Call a meta-function $\mathtt{\mathbf{update}}$, to (atomically) store
  $\mathtt{\mathbf{\vec{t}}}$ in $\mathtt{\mathbf{State}_X[\vec{p}]}$.
\end{itemize}

This program maintains, the following invariant
\begin{propal} \label{prop:inv} If the value of
  $\mathtt{\mathbf{State_X[\vec{p}]}}$ is $\mathtt{\mathbf{\vec{t}}}$, then
  $\mathtt{\mathbf{Tile_X (\vec{p},\vec{t})}}$ has been successfully executed.
\end{propal}

Note that our scheme has none of the overheads of a typical run-time system:
worker queues, schedulers, shared data structures, cactus stacks, etc.  
Not even a mutex---all the writes to the $\mathtt{\mathbf{State}}$ arrays are
guaranteed to be exclusive write.
Key scheduling decisions are made directly in the control structure of the
generated code (hence the term ``self-scheduling'') and this is why we see
significant performance gains.

\subsection{Implementing the Acquire Function}
It follows from Props.~\ref{prop:dep} and~\ref{prop:inv} that all that acquire
must ensure is that for each input-output pair, $\langle {\cal D}, X, Y, [\pi,
\tau] \rangle$ whose source node is $X$,
\begin{itemize}\itemsep 0mm
\item First, test if the current space-time coordinates, $\langle\vec{p}, \vec
  {t} \rangle$ belong to ${\cal D}$
\item If so, ensure that (wait until) the $\pi(\vec{p}, \vec {t})$-th entry in
  $\mathtt{\mathbf{State_Y}}$ is lexicographically greater than or equal to
  $\tau(\vec{p}, \vec {t})$
\end{itemize}

Note how this is completely target agnostic.  In fact, our implementation uses
a simple busy wait.  The cost of the implementation is a finite number (per
input-output pair of the RPRDG) of evaluations of simple affine functions of
the current space-time coordinates, which are about as complicated as affine
address or loop bound calculations in a typical polyhedral program, and it is
done at the granularity of a tile.  One may think that a busy wait may be
expensive, but this is mitigated by the fact that all tiles of a polyhedral
program are equal volume (except boundary ones), and the dependence structure
is regular.

\subsection{Avoiding Deadlock}


If we implement the \texttt{\textbf{acquire}} as a simple busy wait may lead
to deadlock: a process executing a busy wait uses processor resources, and
this may prevent the (virtual) processor that is suposed to execute the
``producer tile'' from advancing.  Because the programs we deal with are
polyhedral, there is a simple way to avoid deadlock.  It requires two
elements.

\begin{itemize}
\item An additional legality condition on the HSD Schedule: for every
  input-output pair $\langle X, Y, {\cal D}, f\rangle$ in the RPRDG, the
  following condition must hold.
  \[\forall z\in{\cal D}, \pi_X(z) \succeq \pi_Y(f(z))\]
  Informally, this states that the producer-consumer mapping among processors
  is in a ``lexicographically increasing'' direction.
\item Ensure (or develop a run-time mechanism that does so) that when virtual
  processors are mapped to physical resources, it is in the lexicographically
  increasing order.  This ensures that once a processor starts, a
  non-preemptive scheduler will not deadlock.
\end{itemize}

\section{Code Generation}
\label{sec:codegen}

This section first describes how to modify the PRDG $\cal H$ to insert
\texttt{\textbf{acquire}} and \texttt{\textbf{update}} statements. Then
presents the code structure and challenges in implementations for different
target platforms/runtimes (i.e. Pthreads, GPU/CUDA and X10).

\begin{algorithm}[h]
  \label{alg:hybrid_algo}
\KwIn{PRDG $\cal H$, HSD Schedule for each node in
    $\cal H$, Residual PRDG $\cal H'$}
\KwOut{Transformed PRDG with \texttt{\textbf{acquire}} and
\texttt{\textbf{update}} Nodes}
\ForEach {node $X$ of $\cal H'$}{
	add node $X_{acq}$ to $\cal H$ \;
	${\cal D}_{X_{acq}} \leftarrow {\cal D}_X$ \;
	\ForEach {input-output-pair $\langle X,*,{\cal D}_e,f \rangle$}{
		\ForEach {pair with same $D_e$}{
			$args \leftarrow args ~\cup ~$ check $(f\left(z\right) $ as a tuple$)$ \;
		}
		$exp \leftarrow D_e : $ acquire $\left(args\right)$ \;	
		$rexp \leftarrow rexp \cup exp$ \;	
	}
	${\cal S}_{X_{acq}} \leftarrow $ XAcquire $ = rexp$ \;
	set the schedule of ${\cal S}_{X_{acq}}$ to execute as the 1st statement in
the tile. \;
}
add node $U$ to $\cal H$ \;
${\cal D}_U \leftarrow \bigcup_{i} {\cal D}_{X_i}$ \;
${\cal S}_U \leftarrow $ update ($z$) \;
set the schedule of ${\cal S}_U$ to execute as the last statement in the
tile \;

\caption{Update the PRDG $\cal H$ with new nodes for
\texttt{\textbf{acquire}} and \texttt{\textbf{update}} statements. In line 8,
$D_e$ : $expression$ means that $expression$ is defined over the domain $D_e$}

\end{algorithm}

Algorithm~\ref{alg:hybrid_algo} describes the steps in modifying the PRDG
$\cal H$ to add \texttt{\textbf{acquire}} and \texttt{\textbf{update}}
statements. We take PRDG $\cal H$, residual PRDG $\cal H'$ and HSD schedule as
inputs. For each node in the residual PRDG $\cal H'$ we add a new node to
$\cal H$. These nodes corresponds to the \texttt{\textbf{acquire}} statements.
The domain of the node is same as the domain of the corresponding node in
$\cal H'$. The schedule of the node is set such that it get executed as the
first statement of the tile. The left hand side of the statement (XAcquire) is
a dummy scalar variable.  The right hand side of the statement (\emph{rexp})
is a union of \texttt{\textbf{acquire}} function expressions, each with domain
$D_e$ -- domain of corresponding edge in $\cal H'$.  \texttt{\textbf{acquire}}
itself is a dummy function which takes dependences as arguments. These
dependence functions are expressed as \texttt{\textbf{check}} functions with
$f(z)$ as argument. $f(z)$ is a $(n)$-tuple of integers where first $k$
integers corresponds to the virtual processor coordinate $\vec{p}$ and rest
corresponds to the time step $\vec{t}$ on which the current tile depends. For
each target platform, \texttt{\textbf{check}} function must be implemented in
such a way that it returns only when the $\vec{p}$ reaches the time $\vec{t}$.
For the \texttt{\textbf{update}} statement, we add only one node to $\cal H$
and we set the domain as the union of domains of all the nodes in $\cal
H$. The schedule is set to execute it as the last statement in the
tile. Memory allocation is $z \rightarrow \pi(z)$ where the dimension of
memory is $k$ -- number of processor dimensions. For rest of the nodes in the
$\cal H$ the schedule is updated so that they get executed in between
\texttt{\textbf{acquire}} and \texttt{\textbf{update}} statements.

For the REX~\ref{fig:prdg}, we add 2 nodes for \texttt{\textbf{acquire}}
statements and one for \texttt{\textbf{update}} statement to the PRDG.
Statement corresponds to edge $e1$ in RPRDG is $\texttt{\textbf{XAcquire}}_1 =
$ \texttt{\textbf{acquire}}(\texttt{\textbf{check}}($i_b-1,j_b$)) where the
domain of the expression $\{i_b,j_b|i_b > 0 \land j_b > 0\}$. The statement
for $e2$ is same except for the domain of the expression. The update statement
is \texttt{\textbf{update}}($i_b,j_b$) with domain $\{i_b,j_b ~|~ 0\leq i_b
\leq M_b \land 0\leq j_b\leq N_b \}$

The final step in code generation involves implementation of
\texttt{\textbf{acquire}} and \texttt{\textbf{update}} functions for different
target platforms.
  
Each target has its own language specific synchronization constructs.  In
order to show the expressiveness of our technique for a variety of targets, we
provide target specific implementation for Pthreads here.  The outline and
issues for implementations in CUDA and X10 are described in an appendix that
has been omitted due to space constraints.

\subsection{Pthreads}

\label{pthreads}

Our Pthreads implementation relies on pthread mutex operations to control
access to the queue of Blocks, and a blend of busy-waiting and mutex
operations to implement of CHECK and UPDATE as each block progresses through
tiles.

Figure~\ref{fig:pThreadQueue} shows the code for the work claiming action to
be performed by each thread to declare the ownership of a block of work.  As
this is used infrequently (once per block start), we expect the overall
performance impact of pthread mutex operations to be negligible.

\lstset{language=C, captionpos=b, caption={\small{pthread code for block claiming}}, label=fig:pThreadQueue}
{
\small
\begin{lstlisting}
void* Process_block() {
  int t1,t2;
  //mutexptr guards access to task_ptr
  //task_ptr always gives lex. minimum
  //  unclaimed block
  pthread_mutex_lock (&mutexptr);
  t1 = __Queue[task_ptr];
  task_ptr++;
  pthread_mutex_unlock (&mutexptr);
  // this process now executes tiles,
  // possibly waiting at tile entry
  for (t2=0; t2<nTiles(t1); t2++)
      Tile(t1, t2);
}
\end{lstlisting}
}

Figures~\ref{fig:pThreadAcquire} and~\ref{fig:pThreadUpdate} show the code run
upon entrance and exit to tile~t of a processor~p As this code is executed
(repeatedly, in the case of UPDATE) for each tile, overhead must be kept low.
Busy-waiting is an appealing option, at least in the frequent case of
rectangular iteration spaces with uniform dependences, as (in such cases) a
thread may arrive at the start of a tile at about the time the sources have
completed, and furthermore we hope to minimize the number of times a thread
switches away from its current block.

While the busy wait can reduce the energy cost of unnecessary switching among
threads, lengthy busy waits can waste both time and energy.  Thus, we perform
a fixed number of iterations of busy waiting (2, as in
Figure~\ref{fig:pThreadAcquire}), after which we resort to pthread mutex
operations.

\lstset{language=C, caption=\small{pthread code for CHECK}, label=fig:pThreadAcquire}
{
\small
\begin{lstlisting}
// called once for each dependent
// tile coordinates (p', t')
//  before execution of any code
//  in tile (p,t)
int check(int p, int t) {
  int current_step = -1;
  int _counter = 0;
  while(current_step < t) {
    //mutex ensures exclusive access
    //to timestamp array
    pthread_mutex_lock (&mutexsync);
    current_step = __STATUS_[p];
    _counter ++;
    // limit on busy waiting
    if (_counter>2) 
      pthread_cond_wait(&sync_cv, &mutexsync);
    pthread_mutex_unlock (&mutexsync);
  }
}
\end{lstlisting}
}

\lstset{language=C, caption=\small{pthread code for UPDATE}, label=fig:pThreadUpdate}
{
\small
\begin{lstlisting}
// Upon exit from time step t inside
// a processor p
int update(int p, int t) {
  //mutex ensures exclusive access
  //to timestamp array
  pthread_mutex_lock (&mutexsync);
  __STATUS_[b]=t;
  pthread_cond_broadcast(&sync_cv);
  pthread_mutex_unlock (&mutexsync);
}
\end{lstlisting}
}

\subsection{CUDA}
\label{cuda}

This section describes the CUDA code structure for REX and challenges of a
GPGPU implementation.  For the CUDA implementation of the hybrid schedules, we
introduced several changes compared to the classic wavefront parallel
implementation.  In wavefront parallel code, there is a kernel call per each
wavefront but for the hybrid schedule there is only one kernel call with
$\frac{M}{b}$ thread-blocks in a 1D grid ($b$ is the size of the tile along
processor dimension).  Each thread-block responsible for the execution of one
column of tiles.  The code structure of the CUDA kernel is shown in
Listing~\ref{lst:cuda_kernel}.

The \texttt{\textbf{update}} and \texttt{\textbf{check}} functions are
executed only by the first thread of the thread-block and all the other
threads wait on a \emph{syncthreads()}.  The \texttt{\textbf{update}} function
simply set the value of array at processor index to the time index of the tile
(see Listing~\ref{lst:cuda_update}). The \texttt{\textbf{check}} function is a
busy wait until the value of array at processor index becomes the time index
of the tile that depends on (see Listing~\ref{lst:cuda_check}).

There are CUDA specific issues that need solutions for this method to work.
The variables that are accessed by multiple Streaming Multiprocessors (SMs)
must be accessed in a way that it skips the L1 cache which is private to the
SM. This can be achieved by skipping L1 for the whole kernel by specifying a
compiler flag or for each variable individually.

The main challenge is avoiding deadlocks caused by the synchronization
mechanism combined with the scheduling mechanism of CUDA thread-blocks. The
execution order of thread-blocks is undefined. If we statically assign the 1st
column of tiles to the 1st thread-block, 2nd column to the 2nd thread-block
and so on, then we assume that the thread-blocks are scheduled in ascending
order which contradicts with CUDA programming semantics.  If the last set of
thread-blocks get scheduled first, CUDA kernel deadlocks since all the thread
blocks are waiting for previous thread-blocks to finish the depending tiles
but there is no resources to schedule these thread-blocks. Hence, we need to
allocate the column number dynamically to the thread-block at run-time.  We
use a simple mechanism proposed by Yan et al.~\cite{Yan2013PPoPP} based on
atomic increment function in CUDA. The first computation of the thread-block
is to increment a variable in global memory using first thread and use the
returned value as the tile column to process (see
Listing~\ref{lst:cuda_next}).  Irrespective of which thread-block get
scheduled next, it will pickup the next available tile column to process.

\lstset{language=C, caption=\small{CUDA code structure for REX-HSD},
label=lst:cuda_kernel}
{
\small
\begin{lstlisting}
//pick the next available column of tiles
if (threadIdx == 0)
  p = getNextColumnofTiles();
syncthreads();
Status[0] = 0;
// T time iterates over tiles
for (t in 0..(T-1)) {
  // ACQUIRE / CHECK state
  if (threadIdx == 0)
    acquire(check(p-1, t));
  syncthreads();

  // Execute Tile(p, t);
	
  // UPDATE status
  if (threadIdx == 0)
    update(p, t); 
  syncthreads();
}
\end{lstlisting}
}

\lstset{language=C, caption=\small{CUDA code structure check function},
label=lst:cuda_check}
{\small
\begin{lstlisting}
check(p,t) {
 while (Status[p] < t) {;}
}
\end{lstlisting}
}

\lstset{language=C, caption=\small{CUDA code structure update function},
label=lst:cuda_update}
{\small
\begin{lstlisting}
update(p,t) {
  Status[p] = t;
}
\end{lstlisting}
}

\lstset{language=C, caption=\small{CUDA code structure to get the next available column id of tiles},
label=lst:cuda_next}
{\small
\begin{lstlisting}
getNextColumnofTiles(p,t) {
  return atomicInc(&columnCount);	
}
\end{lstlisting}
}

\subsection{X10 Code Generation}
\label{x10}

As with the other target platforms/run-time systems, we can implement the
\texttt{\textbf{Acquire/State}} protocol for the X10 language~\cite{X10}.
Listing~\ref{lst:X10code} shows X10 code structure.  We spawn new
``activities'' in X10, that correspond to asynchronously executing threads.
The \textbf{when} clause does a wait until the \texttt{\textbf{State}}
variable has a value of the current time stamp \textit{t}.  However, the
\texttt{\textbf{wait}} construct in X10, does not cause a busy wait, rather
the X10 run-time system manages waiting `activities, and liberates resources
for other activities.  Hence resources are available for, and X10 guarantees
that there will be no deadlock ``by construction.''  So no special deadlock
avoidance mechanism is needed.  As soon as \textbf{\texttt{State}} is updated
by the producing process, all activities waiting on its value are woken up and
placed in the ready queue by the run-time system.

\lstset
{caption=\small{X10 code structure for Rex-HSD},label=lst:X10code}
{\small
\begin{lstlisting}
Status[0] = 1; // base case
// P processors
for (p in 0..(P-1)) 
  // spawn an activity per processor
  async {
  // T time 
    for (t in 0..(T-1)) 
      // ACQUIRE / CHECK state
        when( Status[p-1] >= t ) {
          // Execute Tile(p,t)

          Status[p] = t; // UPDATE
	}
\end{lstlisting}
}

\section{Asymptotic Overhead}
\label{sec:complexity}

We now quantify three main overheads of our system (number of processes,
number of synchronizations, and memory overhead) and show that they are
better, asymptotically, than those of previously proposed methods.  Recall
that we are working with a program representation at the tile level, in the
form of a tile PRDG, and that our schedules are $n$-dimensional, which
corresponds to the maximum number of dimensions in the original computation.
Each PRDG node represents a specific ``tile signature'' and is associated with
a domain that enumerates the set of tile instances with that signature.  These
domains are polyhedral sets, parameterized by one or more size parameters.
For simplicity of the presentation here, we assume that there is a single size
parameter, $S$.  Hence the asymptotic ``work'' complexity of the original
program in $\Theta(S^n)$, a degree-$n$ polynomial.

If we use fixed size tiling, the number of tiles remains $\Theta(S^n)$
asymptotically.  If we use mono-parametric tiling using a single tile size
parameter, $b$, and if we set $M=\frac{S}{b}$, it is $\Theta(M^n)$.  Let us
first discuss fixed size tiling.

\begin{lemma}
  The total number of dependence instances is $\Theta(S^n)$.

  The total number of dependences that are statically [c.f.\ dynamically]
  satisfied is $\Theta(S^n)$.
\end{lemma}

The proof follows from the fact that (i) every instance of a tile has a
constant number of other tiles on which it depends, (ii) in the worst case
all/none of the the dependences are statically satisfied, and (iii) tile shave
a constant size.

\paragraph{Synchronization Overhead}

The immediate conclusion of this is that the overheads of any synchronization
mechanism is as large as a constant factor of the total execution time.  Note
that the mere ``constant factor,'' by which tiling mitigates this can be
relatively high (e.g., many tens of thousands if the tile is $16\times
32\times 256$, a typical value).

As a result, the total number of run-time checks that the generated code will
perform is, asymptotically the same regardless of what specific scheme is used,
provided the proposed scheme is efficiently implemented.  However, the
specific choice of the schedule, as well as the nature of the underlying
synchronization mechanism, may affect the constant factors.

This also implies that all proposed schemes, if implemented in the best
possible way, will be only within a constant factor of each other.   

Mono-parametric tiling provides us with an additional tunable parameter, the
tile size $b$.  As a result, the overhead is no longer a \emph{constant}
fraction of the total computation of the program, but rather, polynomial in
$M=\frac{S}{b}$ of exactly the same degree.  Also note that even with
mono-parametric tiling, it is unlikely that $b$ will grow
\emph{asymptotically} at the same rate as $S$.

\paragraph{Memory Overhead}

We now address the memory overhead of our proposed scheme.  Note that the
overhead corresponds to the memory allocated to the auxiliary variable
\textbf{\texttt{State}} as introduced in Section~\ref{sec:hsd}.  Recall that
each process maintains its ``current state'' as an $(n-k)$-tuple of integers,
representing the most recent local time stamp it has completed.  We use a shared array to maintain this.

\begin{lemma}
  The memory overhead of the HSD schedule is $\Theta(S^k)$, a polynomial whose
  degree is the number of processor dimensions of the schedule.
\end{lemma}

The proof is straightforward since $\Theta(1)$ space is required for each
virtual processor, and the union of the images of the domains of all the nodes
in the PRDG yields a $k$-dimensional union of polyhedra, parameterized by the
size parameter, $S$.  The number of integer points in such a polyhedron is a
$k$-degree polynomial.

And finally, note that as mentioned earlier, all previous
techniques~\cite{dathathri-etal-topc2016, DAGuE, KPPGCS2015DataflowTiled,
  BBDHD2012EuroPar, Baskaran:2009:CDS:1504176.1504209}

\paragraph{Comparison with other techniques}
We now compare the overhead of our approach with other techniques proposed in
the literature.  As stated above, all reasonable approaches have the same
asymptotic complexity in terms of the number of synchronization, it's only in
the memory overhead that they may differ.
\begin{itemize}
\item Kong et al.~\cite{KPPGCS2015DataflowTiled} propose a technique that
  compiles tiled polyhedral programs to the OpenStream dataflow
  lanuage~\cite{OpenStream}.  In their approach, there is a ``flag'' that
  takes up one byte for every synchronization between tasks.  Hence the memory
  overhead of their approach is $\Theta(S^n)$.  Dathatri et
  al.~\cite{dathathri-etal-topc2016} present a similar technique that targets
  distributed memory systems and therefore also handles communication.  Theire
  performance overheads are similar.
\item Belviranli et al.~\cite{belviranli-ics15} propose a similar scheme for
  GPUs, albeit fir a limited class of programs.  They too suffer from an
  overhead of is $\Theta(S^n)$.
\item Boslica et al.~\cite{BBDHD2012EuroPar} propose a scheme to compile
  affine loops to the DAGuE run-time system.  The overhead arises from an
  interplay between their compile-time transformation and the DAGuE run-time.
  If the {\tt Insert\_Task} calls of their Figure~1 corresponds to the
  creation of a DAGuE ``micro-task'', and these have a representation in DAGuE
  of size $> 0$, and the micro-task creation does not itself wait for any
  executing micro-tasks, this produces a memory overhead of $\Theta(S^n)$.
  %
  %
\end{itemize}

Table~\ref{tab:asymptotic_comparison} compares the asymptotic comlexity of
other approaches~\cite{KPPGCS2015DataflowTiled, dathathri-etal-topc2016,
  DAGuE, belviranli-ics15} with ours for a number of benchmarks that we later
use in our exprimental validation: Jacobi-1D (J1D) Jacobi-1D (J1D) and Wave-3D
(W3D) are all stencils, REX3D is a 3D iteration space version of REX, Chol
Cholesky decomposition (Chol) and Lower Triangular Matrix Inversion (LTMI) are
dense linear algebra, and two versions of Optimal String Parenthesization (OSP
and OSPGKT) are dynamic programming.  The last letter after the benchmarks
indicate whether fixed (F) or CART (C) tiling is used.

\begin{table}
\scalebox{0.76}{
  \centering\footnotesize
\begin{tabular}{|l|c|c|c|c|c|c|c|c|c|c|c||l}
  \hline
  \multicolumn{3}{ | c | }{\textbf{Benchmark}}  & \multicolumn{8}{ | c |
  }{\textbf{Approach}} \\ \hline 
  \textbf{Name} & \textbf{Work} & \textbf{\#Tiles} & \multicolumn{2}{ | c
    | }{\cite{dathathri-etal-topc2016} / \cite{KPPGCS2015DataflowTiled}} &
  \multicolumn{2}{ | c | }{\cite{DAGuE}} &
  \multicolumn{2}{ | c | }{\cite{belviranli-ics15}} & \multicolumn{2}{ | c |
  }{HSD}\T \B \\ 
  \hline 
 & & & \#T & \#S & \#T & \#S & \#T & \#S & \#T & \#S \T \\ \hline 
 
J1D\_F &$N T$ & $\frac{NT}{b^{2}}$ & 1 & 3 / 2 & - & - & $\frac{N}{b}$
& 1 &$\frac{N}{b}$, $\frac{T}{b}$ & 1 \T \\ \hline
 
J2D\_F &$N^2 T$ & $\frac{N^{2}T}{b^{3}}$ & 1 & 7 / 3 & - & - &
* & * & $\left(\frac{N}{b}\right)^2$, $\frac{N}{b}$ &1,3 \T \\
\hline

W3D\_F &$N^3 T$& $\frac{N^{3}T}{b^{4}}$ & 1 & & - & - & & &
$\frac{b^{3}}{N^{3}}$ & 1 \T \\ \hline

REX3D\_C &$N^3$ & $\frac{N^{3}}{b^{3}}$ & 1 & 3 & - & - & - & -
&$\left(\frac{N}{b}\right)^2$, $\frac{N}{b}$ & 1,2 \T \\ \hline

Chol\_C &$\frac{N^3}{6}$ & $\frac{N^{3}}{6b^{3}}$ & $1$ & 3 & $1$ & 3
& - & - &$\left(\frac{N}{b}\right)^2$ & 1 \T \\ \hline

LTMI\_C &$N^3$ & $\frac{N^{3}}{b^{3}}$ & 1 & 3 & 1 & 3 & - & - &
$\left(\frac{N}{b}\right)^2$ & 1 \T \\ \hline


GKT\_C &$N^3$ & $\frac{N^{3}}{b^{3}}$ & 1 & 3 & - & - & - & - &
$\left(\frac{N}{b}\right)^2$ & 1 \T \\ \hline
\end{tabular}
}
\caption{\small{Comparison of Asymptotic Overhead of the number of tiles per
    task (\#T) and number of synchronizations per tile (\#S) of other methods
    with ours for a number of benchmarks.  For J2D\_F, Belviranli et
    al.~\cite{belviranli-ics15} have a factor of $b$ more tiles than all other
    methods, so their numbers are not reported.} }
\label{tab:asymptotic_comparison}
\end{table}

\section{Experimental Evaluation}
\label{sec:experiments}

Hybrid Static/Dynamic Scheduling algorithm is applicable to all Polyhedral programs.  We conduct experiments on different types of polyhedral programs. Our experiments include benchmarks from Dynamic programming, Linear Algebra and Stencil class of programs. To compare with performance of the same kernels generated by state-of-the-art compilers, we would like to compare our generated code with both static and dynamic schedules. However, due to the difficulties involved in acquiring some of the compilers, we were only able to compare our performance with statically scheduled code with OpenMP generated by 2 polyhedral compilers PLuTo and DTiler, which are known to achieve high performance.

Our benchmarks include REX3D, which is a dynamic programming algorithm;
Lower Triangular Matrix Inversion (LTMI) and Cholesky Decomposition, which are linear algebra methods;
and Jacobi\_1D as well as Jacobi\_2D, which are stencils.
In our experiments, Hybrid Static/Dynamic Schedules are applied to two classes of programs,
producing two variants of HSD scheduled codes: mono-parametrically tiled programs that we obtained by instrumentation on CART generated codes,
and programs with constant tile sizes obtained by instrumentation on PLuTo generated codes. The instrumentation of both kernels is done by inserting point-wise synchronization functions whose domain is derived using existing polyhedral tools such as iscc followed by the removal of previously generated OpenMP pragmas. For practical reasons, we focus on deriving stencil codes with HSD schedules from PLuTo generated code and dynamic programming and linear algebra methods with HSD schedules from CART generated codes.
For each of these benchmarks, we choose a target mapping.
The polyhedral code generator gives us the flexibility to use any legal target mapping.
Therefore, we explore two target mappings each for REX\_3D and Jacobi\_2D. Experimental target mappings are shown in Table~\ref{tab:benchmark_mappings}.


\begin{table}

\scalebox{0.95}{

\centering\footnotesize
\begin{tabular}{|l|r|r|r||l}
\hline
 \textbf{Benchmark \&} & \textbf{Target Mapping 1} & \textbf{Target Mapping 2} \T \B \\
 \textbf{Problem size} & & \\
 \hline
 
J1D\_F &  $( t , i ) \rightarrow ( t , 2t+i )$ & \T \\
 $262k\times 2.62M$ & $t$ is processor & \\
 $64k\times 640k$ & $i$ is time  & \\  
 & & \\  \hline
J2D\_F &  $( t , i , j ) \rightarrow$ \hspace*{1.2cm} &  $( t , i , j ) \rightarrow$  \hspace*{1cm} \T \\
 $1k \times 2k \times 2k$ &$( t , 2t+j , 2t+k )$ & $( t , 2t+j , 2t+k )$  \\
 $0.5k \times 1.25k \times 1.25k$ & $t$ is processor & $t , i$ is processor \\
 & $i , j$ is time  & $j$ is time  \\ \hline
 J1D\_C &  $( t , i ) \rightarrow ( t , i )$ & \T \\
 $262k\times 2.62M$ & $i$ is processor & \\
 $64k\times 640k$ & $t$ is time  & \\ 
 & & \\  \hline
 REX3D\_C  &  $( i , j , k ) \rightarrow ( i , j , k )$ & $( i , j , k ) \rightarrow ( i , j , k )$ \T \\
$(2048)^3$ & $i$ is processor & $i , j$ is processor \\
$(2048)^3$& $j , k$ is time  & $k$ is time  \\  \hline
 Chol\_C &  $( i , j , k) \rightarrow ( i , j , k)$ & \T \\
$(2400)^3$ & $i$ is processor & \\
& $j,k$ is time  & \\\hline

 LTMI\_C &  $( i , j , k) \rightarrow ( i , j , k)$ & \T \\
$(2400)^3$ & $i$ is processor & \\
 & $j,k$ is time  & \\\hline

\end{tabular}
}
\caption{\small{Benchmarks and Target Mappings}
\label{tab:benchmark_mappings}}
\end{table}

\begin{table}

\centering\small
\begin{tabular}{|l|r|r|r|l|}
\hline
\textbf{Architecture Parameters }&   &\T \B \\ \hline
Processor &  E3-1231v3 &   E5-2650v2\T \\ \hline
Base Frequency & 3.4 GHz & 2.6 GHz\T \\ \hline
Turbo Boost Frequency & 3.8 GHz & 3.4 GHz\T \\ \hline
Number of Cores &  4 & 8\T \\ \hline
Number of Threads & 8 & 16\T \\ \hline
L3 Cache &  8 MB & 20 MB\T \\ \hline
L2 Cache &  256 KB/core & 256 KB/core\T \\ \hline
L1 Cache &  12 KB/core & 32 KB/core\T \\ \hline
Instruction Set Extensions & AVX 2.0 & AVX\T \\ \hline
Max Memory Bandwidth & 25.6 GB/sec &  59.7 GB/sec\T \\ \hline
Max \# of Memory Channels & 2 & 4\T \\ \hline
\end{tabular}
\caption{\small{Machine configuration}
\label{tab:machine}}
\end{table}

\begin{figure}[t]
  \centering
  \includegraphics[trim = 30mm 70mm 30mm 75mm, scale=0.45]{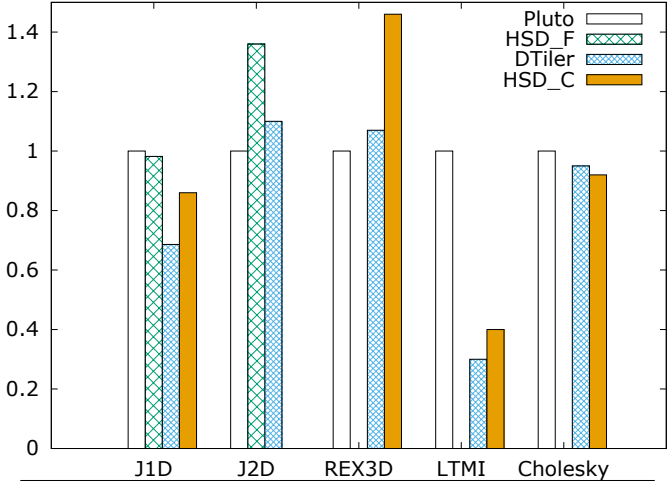}
  \caption{\small{Normalized Gigaflops Achieved on E3-1231V3}}
  \label{fig:perfHaswell}
\end{figure}

\begin{figure}[t]
  \centering
  \includegraphics[trim = 30mm 70mm 30mm 75mm, scale=0.45]{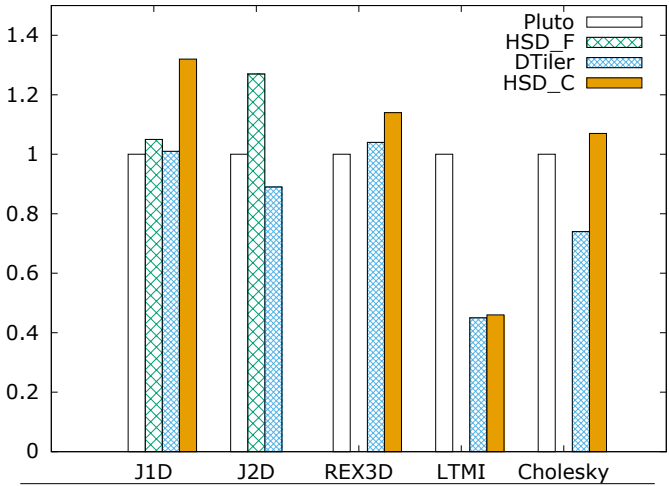}
  \caption{\small{Normalized Gigaflops Achieved on E5-2650V2}}
  \label{fig:perfIvyBridge}
\end{figure}

Experiments are conducted on the Intel Xeon E3-1231-v3 platform, which belongs to the Haswell micro-architecture family,
and the Intel Xeon E5-2650-v2 platform, which belongs to the IvyBridge family.
Detailed machine configuration is shown in Table~\ref{tab:machine}.
We set our problem sizes to have the memory footprint of each benchmark to exceed the last level cache size,
and compile our kernels using icc with the flags \textit{-O3 -xHost -funroll-loops}.
We then perform a search to obtain the best tile sizes for each combination of benchmark, architecture(Haswell, Ivybridge), compiler (PLuTo, DTiler).
We searched in a tile size space of $(1024\times1024)$ for kernels with 2D iteration space and $(256\times256\times256)$ for kernels with 3D iteration space.
The stride of our search is set to 16 in each dimension for parametrically tiled kernels,
and 32 in each dimension for kernels tiled with constant sizes.
The highest achieved  Gigaflop number is normalized with Pluto as the baseline and reported in Figure~\ref{fig:perfHaswell} and Figure~\ref{fig:perfIvyBridge}.
We measured the execution efficiency of the kernel using the best tile sizes discovered and resultant data is shown in Table ~\ref{tab:exec-eff}.
In this table, LLC\_DM indicates the last level cache data misses event count in millions reported directly by running perf~\cite{Perf} to profile specific kernels.
Sync Overhead refers to the percentage of time threads spent waiting for synchronizations to take place over the entire execution time. This is directly quoted from the report generated by Allinea~\cite{Allinea}. Total Energy refers to the total energy consumption of the program measured with unit milliwatt-hour. This figure consists of two parts: CPU and DRAM energy consumption. We measure two data separately. The former is, again, directly quoted from Allinea performance report and the latter is obtained by accessing PowerCap~\cite{PowerCap} interface and measuring energy readings for powerzone named "dram". Both mechanisms gather energy data by utilizing Intel Running Average Power Limit (RAPL) interface.

From Figure~\ref{fig:perfHaswell} and ~\ref{fig:perfIvyBridge},  we observe that except for the case of LTMI, the performance of benchmarks using Hybrid Static/Dynamic Schedules are not significantly behind others using wavefront schedules generated by state-of-the-art polyhedral compilers. If we compare the performance of PLuTo generated code with HSD F and DTiler with HSD C, we can observe that the maximum performance loss was 2\%(J1D) and 4\%(Cholesky) respectively.

Moreover, as evidenced by our execution efficiency data on three of the four
kernels with 3 dimensional iteration space, kernels with Hybrid Static/Dynamic
Schedules are consistently the most energy-efficient ones. On average, our
DTiler and PLuTo counterparts used 1.5x and 1.4x as much energy as HSD kernels
consumed respectively. Therefore in many cases, adopting a Hybrid
Static/Dynamic schedule can improve the energy efficiency of a program at no
cost or very little cost.

LTMI is indeed an outlier in our benchmarks. By examining the vectorization
report we were able to identify that potential unaligned data access in
parametric tiled programs (DTiler and Monoparam alike) can present significant
obstacles to vectorization. However, this is a limitation of our input
program, not a restriction of our run-time mechanism.

Lastly, our runtime scheduling policy mandates that each processor must follow
a strictly lexicographically ascending order within the block and finish all
of the work within a block before being preempted. Such policy essentially
guarantees a multi-pass execution of the iteration space, which was previously proven~\cite{ZR:15} to exhibit energy
efficiency but was only applicable to stencil kernels.

\begin{table}
\centering\small
\scalebox{0.90}{
\begin{tabular}{|l|c|c|c|c|c|c|c|c|l|}
\hline
 \textbf{Kernel} & \textbf{Code} & \multicolumn{2}{|c|}{\textbf{LLC}}  & \multicolumn{2}{|c|}{\textbf{Sync}} & \multicolumn{2}{|c|}{\textbf{Total}}   \T \B \\
 &  & \multicolumn{2}{|c|}{\textbf{DM }} & \multicolumn{2}{|c|}{\textbf{Overhead}} & \multicolumn{2}{|c|}{\textbf{Energy}} \\
  &  & \multicolumn{2}{|c|}{{millions }} & \multicolumn{2}{|c|}{{\%}} & \multicolumn{2}{|c|}{{milli-walthour}} \\
 \hline
  &  & & gain & & gain & & gain \\
 \hline\hline
{J2D} & DTiler & 22 & 3.7x  & 15.10 & 10x & 22.0 & 1.5x \T \\ \hhline{~----}
             & Pluto & 39 & 6.5x   & 19.60 & 13x & 20.2 & 1.4x  \T \\ \hhline{~----}
             & HSD F& 6 &     & 1.50 &  & 14.5 &  \T \\ \hline \hline
{REX 3D}       & DTiler  & 30 & 4.3x  & 4 & -0.72x & 51.0 & 1.5x  \T \\ \hhline{~----}
            & Pluto & 7 & 1x     & 11.90 & 2.2x & 58.7 & 1.7x  \T \\ \hhline{~----}
            & HSD C& 7 &     & 5.50 &  & 34.8 &  \T \\ \hline \hline
{Cholesky}  & DTiler & 16 & 5.3x   & 8.50 & 1x & 15.7 & 1.6x  \T \\ \hhline{~----}
            & Pluto & 3 & 1x     & 28.60 & 3.4x & 12.1 & 1.24x  \T \\ \hhline{~----}
            & HSD C& 3 &      & 8.30 & & 9.7 &  \T \\ \hline

\end{tabular}
}
\caption{\small{Execution Efficiency}
\label{tab:exec-eff}}
\end{table}

\section{Related Work}
\label{relatedwork}

The combination of automatic loop transformation and data-flow run-time
synchronization has been explored on a number of occasions.  Our work is novel
in several respects; here, first we list the novel features of our approach
and then discuss specific other systems.  

Motivation for, and experimental validation of, prior data-flow run-time
synchronization systems has focused primarily on execution speed, typically
framed in terms of issues of idle processors and load balance, rather than
energy use.  Note that cache misses impact both speed and energy use.
Although it may be possible to hide (via prefetching) the \emph{delay} of a
cache miss, it is not possible to offset the \emph{energy consumption}.  Our
approach, like prior work, prevents the unnecessary idling of processors due
to overly strict wavefront synchronization.  Unlike other work, our approach
can ensure locality statically, without reliance on luck or run-time locality
analysis, thanks to our specification of process coordinates.

Our use of processor coordinates also provides an unusually tight coupling
between compile-time transformation/scheduling and run-time scheduling and
synchronization, resulting in scalability and overhead advantages for our
implementation.  By using one thread \emph{per slice of tiles} rather than
\emph{per tile}, we reduce the number of threads, thus demanding less storage
for thread contexts and performing fewer thread-switches.  For synchronization
overhead we use one integer \emph{per slice of tiles}, rather than one Boolean
value \emph{per tile}.

Bosilca et al.~\cite{BBDHD2012EuroPar} present a static analysis to derive the
dataflow patterns of nested C loops that use calls to PLASMA library
routines~\cite{PLASMA-UserGuide} to perform ``tile-based dense linear algebra
algorithms.''  They use this inter-tile dataflow information to construct a
set of tasks to be executed by the DAGuE run-time system~\cite{DAGuE}.  DAGuE
uses data-flow run-time synchronization, and employs a system of local and
global queues in the run-time system, organized to ``favor the cache reuse
effect''~\cite[Section~3.2]{DAGuE}.  They also report execution speeds that
beat ScaLAPACK and equal hand-tuned codes on a distributed system based on
Xeon chips with with a total of 648 cores.  They do not discuss energy
consumption.  While this study focuses on a manycore system, a port of the
PLASMA library and the DAGuE run-time system would provide a broad
multi-target system.  We believe that the dataflow among tiles of our tiled
PRDG from Section~\ref{sec:hprdg} is essentially similar to the inter-tile
dataflow collected by Bosilca et at., but note that our system is not
restricted to a library of specific tile operations for square
tiles~\cite[Section~2.1.1]{PLASMA-UserGuide}.  In our approach affinity
between tiles of the same block arises automatically, rather than relying on a
run-time system, and (as discussed earlier) our approach spawns asymptotically
fewer threads.

Kong et al.~\cite{KPPGCS2015DataflowTiled} describe an algorithm that uses the
polyhedral model to compile C loop nests into the task-parallel dataflow
language OpenStream~\cite{OpenStream}.  They motivate their work in terms of
locality improvement between subsequent loops that manipulate the same array,
rather than between the tiles of a single loop nest, and demonstrate that it
can produce much lower memory traffic and execution time than two different
approaches to loop fusion (as well as performance results for codes that
require tiling).  Like Bosilca et al., they do not measure energy consumption,
and do not provide a static mechanism to ensure inter-tile locality (they note
that affinity information could be derived from their representation, but
``the current OpenStream implementation we used does not allow provision of
scheduling guidelines and therefore locality opportunities were often lost due
to the task firing policy of
OpenStream''~\cite[Section~5]{KPPGCS2015DataflowTiled}).

The PeerWave approach proposed by Belviranli et al.~\cite{belviranli-ics15}
uses data-flow runtime synchronization for GPGPU code that executes tiled loop
nests.  The approach was implemented only for GPGPU's, and experimental
results focus primarily on run-time speed, and the authors also noted a minor
improvement in L2 cache misses
but discuss its relevance to execution speed rather than energy consumption.
Their synchronization code~\cite[Algorithm~2]{belviranli-ics15} uses one
Boolean value per tile, rather than one integers per tile thread, leading to
the potential scaling issues discussed above.  Moreover, they only implemented
the core of their their approach for 2D iteration spaces, and handle higher
dimensional domains by a sequence of PeerWave parallelizations of just the
inner two loops.  This causes them to lose significant benefits of tiling and
locality.

One important special case of our analysis is the absence of \emph{any}
inter-block dataflow, i.e., ``synchronization-free parallelism.''  In this
situation, all blocks can execute concurrently, each on its own processor.
Techniques for uncovering schedules with synchronization-free parallelism date
back at least to the work of Lim and Lam~\cite{LimLam1997}, and was extended
to handle non-affine cases by Beletska et
al.~\cite{BBCP2009SyncFreeNonAffine}.  Our contribution focuses on issues of
synchronization and locality, not scheduling, and is orthogonal to this and
other scheduling work.

In addition to the aforementioned implemented systems, some early work
suggested the use of dataflow-based synchronization without providing an
actual implementation.  Kelly and Pugh~\cite{KP:96} discussed the possibility
of moving barriers further out within a nest of loops, and using
``post-and-wait style synchronization'' as part of their mechanism for
estimating communication cost.  They did not, however, investigate actual code
generation for such a system.  Similarly, Wonnacott~\cite{Wonnacott:00}
suggested that tiles might start as soon as their input data was ready, but
did not provide an implementation.

Alias and Plesco~\cite{AliasPatent} proposed a similar process in the context
of high-level synthesis of process networks where processes are linked through
communication channel.  Each channel is associated with a synchronization unit
which might decide to freeze a producer or a consumer process if it is too
advanced in its computation and might start violating the legality of the
program.

Our simple idea of deadlock avoidance via lexico-positive communication in the
processor space is based on early ideas from systolic array synthesis.
Derrien et al.~\cite{sanjay-isss01} showed that the lexico-positive
interconnections were a necessary condition for Locally Parallel, Globally
Sequential (LPGS) partitioning~\cite{fortes-moldovan}.  Greibl~\cite{GLW98}
proposed a scheme called ``forward communication only'' for an early
polyhedral code generator, that imposed such a condition, although it was not
strictly necessary for their codes.

A preliminary version of this work was presented as a poster; details are
omitted for the double-blind review.

\section{Conclusion}
\label{conclusion}

Traditionally the polyhedral model and many techniques at the foundation of
languages and compilation have exploited the fact that information available
at compile time, i.e., \emph{static} analysis, would lead to efficient code.
Recently, this idea has been challenged by the argument since machines are so
incredibly complex, and therefore static techniques alone are not going to
yield good performance, as evidenced by the popularity of auto-tuning, and
data-flow techniques, even for \emph{highly regular}, e.g., polyhedral
programs.

The traditional wavefront approach to scheduling and synchronizing the
execution of tiled loop nests has a number of significant drawbacks that have
been observed by many authors.  The overly strict synchronization semantics
can cause needless idling of processors.  It can force processors to flush
useful data from cache to switch to another tile in the same wavefront.  And
the semantics of barrier synchronization do not fit some target
infrastructures (e.g., CUDA) naturally.

We have shown that these problems can be resolved, while retaining the
benefits of the polyhedral loop transformation framework, by (i) making the
virtual processor coordinates an explicit part of the mapping.  Within this
single framework, we can generate code for diverse targets, allowing each to
use appropriate mechanisms to address issues of synchronization mechanism,
deadlock avoidance, data affinity, etc.  Our performance tests of our
Pthreads-based multi-core code shows that our approach uses less energy than
other systems, either by reducing execution time or by providing competitive
execution time with far better cache performance.

\newpage
\bibliographystyle{abbrvnat}
\bibliography{PPoPP2017}

\end{document}